\documentclass{osa-article}

\journal{oe}

\usepackage{siunitx} 
\usepackage[utf8]{inputenc} 

\begin{document}

\title{Transient vibration imaging with time-resolved synthetic holographic confocal microscopy}

\author{Martin Schnell,\authormark{1,*}P. Scott Carney,\authormark{2} and Rainer Hillenbrand\authormark{3,4}}

\address{\authormark{1}Beckman Institute for Advanced Science and Technology, University of Illinois at Urbana-Champaign, Illinois 61801, USA\\
\authormark{2}The Institute of Optics, University of Rochester, Rochester, New York 14627, USA\\
\authormark{3}CIC nanoGUNE and UPV/EHU, 20018 Donostia-San Sebastian, Spain\\
\authormark{4}IKERBARSQUE, Basque Foundation of Science, 48013 Bilbao, Spain}
\email{\authormark{*}schnelloptics@gmail.com}

\newcommand{\pll}{{\mkern3mu\vphantom{\perp}\vrule depth 0pt\mkern2mu\vrule depth 0pt\mkern3mu}} 
\newcommand{\vect}[1]{\ensuremath{\mathbf{#1}}} 



\begin{abstract}
We introduce a new modality for dynamic phase imaging in confocal microscopy based on synthetic optical holography. By temporal demultiplexing of the detector signal into a series of holograms, we record time-resolved phase images directly in the time domain at a bandwidth as determined by the photo detector and digitizer. We demonstrate our method by optical imaging of transient vibrations in an atomic force microscope cantilever with 100 ns time resolution, and observe the dynamic deformation of the cantilever surface after excitation with broadband mechanical pulses. Temporal Fourier transform of a single data set acquired in 4.2 minutes yields frequency and mode profile of all excited out-of-plane vibration modes with sub-picometer vertical sensitivity and sub-micrometer lateral resolution. Our method has the potential for transient and spectroscopic vibration imaging of micromechanical systems at nano- and picosecond scale time resolution.
\end{abstract}

\section{Introduction}
Digital holographic microscopy (DHM) and interferometric microscopy (IM) are enabling techniques in optical microscopy that provide highly sensitive, non-contact surface profiling of micromechanical systems\cite{RN66,RN194}. Dynamic phase imaging enables characterization of motion and response time of movable parts, which yields valuable information for the development of reliable devices and quality screening\cite{RN195,RN194,RN196}. In wide-field modalities, however, the low frame rate of standard camera technology severely limits the resolvable vibration frequency, prompting the development of novel technical solutions. High-speed cameras were shown to enable imaging of out-of-plane vibration modes with kHz and MHz frame rates\cite{RN145, RN161, RN162, RN198}. Alternatively, the speed limitation of conventional camera technology was mitigated by employing stroboscopic illumination schemes, and vibration imaging of microelectromechanical systems (MEMS), surface and bulk acoustic wave devices (SAW, BAW) at up to GHz frequencies was demonstrated\cite{RN143,RN147, RN149,RN156,RN157, RN160, RN163, RN167,RN199}. However, the vertical sensitivity is typically found to be on the tens of picometer to nanometer scale owing to the inferior shot noise of cameras in comparison to photo detectors, although recent developments in camera technology promise improved vertical sensitivity\cite{RN193}. 

In contrast to wide-field imaging modalities, point-scanning microscopy methods such as confocal microscopy naturally provide high-speed, low noise photo detection. Fast photo detectors with response time down to the picosecond range are commercially available. In combination with interferometric methods for phase retrieval, this potential can be leveraged for optical, non-contact vibration imaging at GHz frequencies\cite{RN146, RN148, RN166 }. State-of-the-art methods are based on heterodyne interferometry and routinely achieve an excellent vertical sensitivity on the single-digit femtometer scale\cite{RN196, RN195,RN194, RN205, RN140,RN206, RN164, RN207,RN200,RN204,RN201, RN202, RN213}. However, complete characterization of all out-of-plane vibration modes in a device is typically a time consuming process. This is because (a) the widely employed lock-in demodulation only detects a single vibration frequency at a time and (b) pixel dwell times on the millisecond scale are needed to reach single-digit femtometer vertical sensitivity. Pixel rates are often in the range of 10s to 500 points/s\cite{RN206,RN146,RN148,RN166}, and thus imaging of several vibration modes is usually an hour-long process. Methods based on \emph{direct} recording of the vertical displacement with fast demodulators and subsequent Fourier transform (FT) enable simultaneous vibration detection in a given frequency band and hence much improved imaging times\cite{RN201,RN202}. However, the signal bandwidth is limited by the time constant of the demodulators and ultimately, by the local oscillator (LO) frequency of heterodyne interferometry (usually 35 - 613 MHz)\cite{RN201,RN202,RN213}, placing a lower bound on the time resolution for the imaging of transient vibrations. It is thus desirable to remove the limitation of LOs and demodulators and to directly record the optical phase at up to the time resolution of the photo detector.

Here  we present direct time-resolved phase imaging in confocal microscopy based on synthetic optical holography. We demonstrate it by imaging transient vibrations in an atomic-force microscopy (AFM) cantilever with 100 nsec time resolution after excitation with sinc and chirped pulses. By Fourier analysis, we show that frequency and mode profiles of all excited out-of-plane vibrations can be determined from a single acquisition. 

\section{Method}
\begin{figure}[tb]
		\centering
		\includegraphics[width=0.65\linewidth, trim= 1.1in 5.6in 4.2in 1.6in, clip]{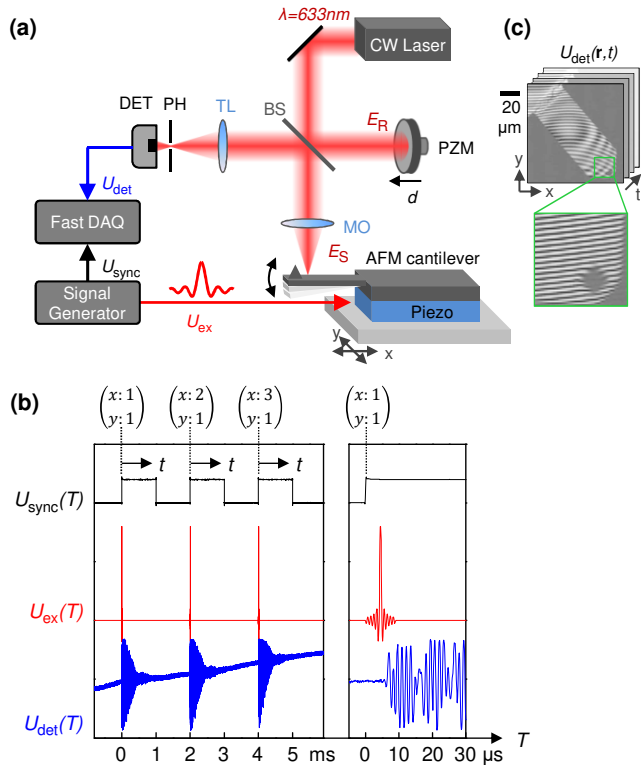}		
		\caption{Time-resolved phase imaging in confocal microscopy with synthetic holographic confocal microscopy. (a) Setup. CW Laser: consisting of a stabilized 2mW HeNe laser, Faraday isolator and beam expander, BS: beam-splitter (50:50 nonpolarizing), MO: microscope objective (20x, 0.4 NA Nikon E Plan), TL: Lens (f = 25.4mm), PH: pin hole (\SI{100}{\micro\meter} diameter), PZM: Piezo-actuated mirror (Physikinstrumente, P-611.3S), DET: photo detector (Thorlabs, PDA36A), DAQ: Data acquisition card (GaGe CS8244). (b) Timing diagram showing the synchronization of the excitation signal $U_\text{ex}$ with the recording of the detector signal $U_\text{det}$ at each pixel $\vect{r}=(x,y)$. Time $t$ marks the time-delay from the beginning of the excitation pulse. (c) 3D Hologram stack $U_\text{det} (\vect{r},t)$ as obtained by demultiplexing the data from (b) after image acquisition, yielding a time series of 20,000 individual holograms.}	
		\label{fig:setup}
	\end{figure}

Fast time-resolved phase imaging with synthetic optical holography (SOH)\cite{RN176, RN175} and its application to transient vibration imaging is illustrated in Fig. \ref{fig:setup}(a). The basis is a confocal microscope setup for optical surface profiling based on SOH\cite{RN175}. The scattered field $E_\text{S}(\mathbf{r})$ from the sample is collected and superposed with a reference field $E_\text{R}(\mathbf{r})=A_\text{R}e^{i\varphi_\text{R}(\vect{r})}$ at the photodetector. The phase of the reference field $\varphi_\text{R}(\vect{r})=4\pi d(\vect{r})/\lambda$ is controlled by the position $d(\vect{r})$ of a piezo-actuated mirror PZM. While raster-scanning the sample, the PZM is slowly translated at a constant velocity to synthesize a linear reference wave $E_\text{R}(\vect{r})=A_\text{R} e^{i\vect{k}_\pll \vect{r}}$, which encodes amplitude and phase of $E_\text{S} (\vect{r})$ in form of a fringe pattern in the fashion of off-axis holography, $U_\text{det}(\vect{r})= |E_\text{S} (\vect{r}) +  E_\text{R} (\vect{r})|^2$. In case of the cantilever sample presented below, the sample stage velocity along the fast axis was $v_x = \SI{199.6}{\micro\meter\per\sec}$ and the PZM velocity was $v_\text{R}= \SI{87.2}{\nano\meter\per\sec}$. With our original setup only stationary objects could be imaged and dynamic phase imaging was not possible\cite{RN175}. For transient vibration imaging, we implemented a high-speed data acquisition scheme where the sample is repeatedly excited to generate vibrations and the amplitude and phase of the scattered field $E_\text{S}(\mathbf{r})$ from the sample is recorded in the time domain at high speed in form of a 3D hologram stack. Each individual hologram is a snapshot of the sample scattered field $E_\text{S}(\mathbf{r})$ at the same time instant, and reconstruction yields the momentary surface profile of the sample. In more detail, the sample is excited with a pre-defined waveform $U_\text{ex}(t)$ at each pixel $\vect{r}$ of the image. Out-of-plane vibrations modulate the phase of the reflected light $E_\text{S}(\mathbf{r})$ and are captured by continuous recording of the detector signal $U_\text{det}(T)$ with a high-speed data acquisition card. Demultiplexing of the detector signal $U_\text{det}(T)$ produces a time-resolved series of synthetic hologram images $U_\text{det}(\mathbf{r},t)$ (Figs. \ref{fig:setup}(b) and \ref{fig:setup}(c)). The momentary complex field $E_\text{S}(\mathbf{r},t)$ at time $t$ can then be reconstructed from individual holograms $U_\text{det}(\mathbf{r},t)$ by spatial filtering in Fourier space
	\begin{equation}
	\label{eq:Udet_tilde}
		\tilde{U}_\text{det}(\vect{q},t) = C(\vect{q},t) + A_\text{R}\tilde{E}_\text{S}^{*}(\vect{k}_\pll - \vect{q},t) +  A_\text{R}^{*}\tilde{E}_\text{S}(\vect{k}_\pll + \vect{q},t) ,
	\end{equation}
where the tilde indicates 2D Fourier transform with respect to \vect{r}. $C$, $A_\text{R}\tilde{E}_\text{S}^{*}$,  $A_\text{R}^{*}\tilde{E}_\text{S}$ are the autocorrelation, conjugate and direct terms, respectively . The momentary surface profile $h(\vect{r},t)$ can subsequently be calculated from the phase $\varphi_\text{S} (\vect{r},t)$ of the complex field $E_\text{S} (\vect{r},t)$ according to
	\begin{equation}
	\label{eq:h}
		h(\vect{r},t) = (\varphi_\text{S}(\vect{r},t) - \varphi_\text{DC}(\vect{r}))\cdot\lambda/4\pi ,
	\end{equation}
where $\lambda=632.8 \text{nm}$ is the wavelength of the illuminating laser beam and subtraction of the time-averaged phase $\varphi_\text{DC}$ from $\varphi_\text{S}$ removes static surface deformation in the sample. 

\section{Transient vibration imaging}

We demonstrated our method by imaging out-of-plane transient vibrations in an AFM cantilever which serves as a well-understood example for micromechanical systems (Fig. \ref{fig:setup}(a)). We excited mechanical vibrations in the cantilever beam by means of a piezoactuator located below the cantilever. The excitation waveform was a broadband pulse (sinc function, $10 V_\text{pp}$ amplitude, spanning a frequency range from 0 to 1 MHz) which was designed to simultaneously excite the fundamental vibration frequency at 46 kHz and higher harmonics in the cantilever (Nanoworld, PNP-DB,  \SI{100}{\micro\meter} long and \SI{40}{\micro\meter} wide). We obtained a 3D hologram stack with 256 x 256 x 20,000 data points in (x,y,t) (Fig. \ref{fig:setup}(c)), where we used a sampling rate of 10 MHz (100 ns time resolution) and chose a recording time of 2 ms per pixel to allow for a complete ring down of all excited vibration modes. The total acquisition time was only 4.2 minutes.  

\begin{figure}[tb]
		\centering
		\includegraphics[width=0.65\linewidth, trim= 1.1in 4.0in 4.2in 1.8in, clip]{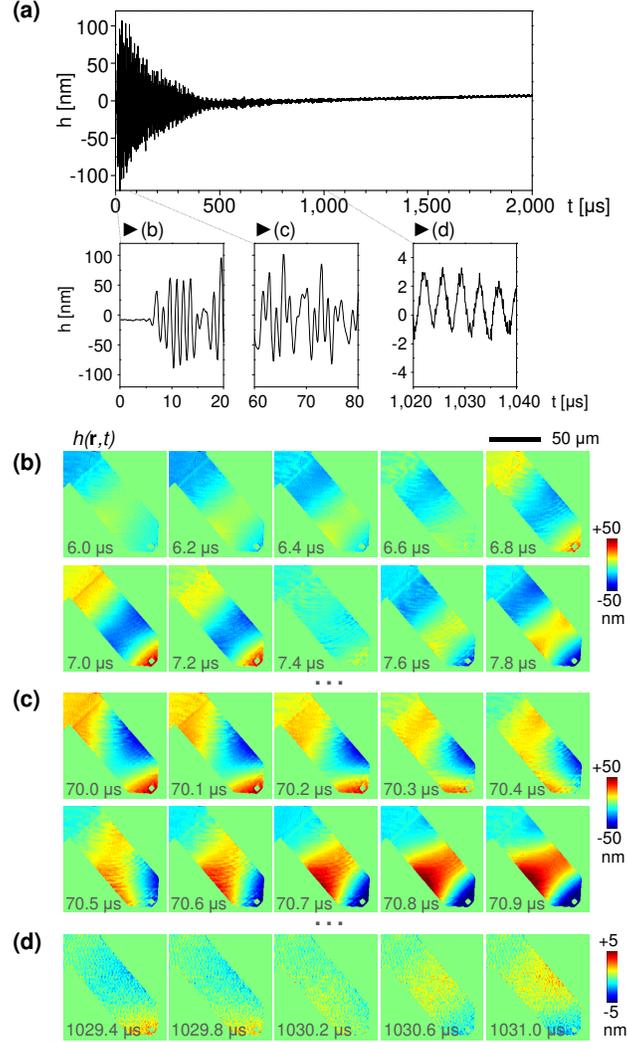}		
		\caption{Imaging of transient vibrations in an AFM cantilever excited with a mechanical broadband pulse with 100 ns time resolution. (a) Time trace of the vertical position $h(t)$ extracted at a pixel near the cantilever tip. Zooms show the time trace immediately after mechanical excitation (b), peak oscillation amplitude (c) and after ring down (d). (b-d) Evolution of the surface profile $h(\vect{r},t)$ of the cantilever (see also \textcolor{urlblue}{Visualization 1}).}	
		\label{fig:time}
	\end{figure}

From the retrieved surface profiles $h(\vect{r},t)$ we exemplarily show the time trace of the vertical oscillation of the cantilever tip, revealing a sharp onset and a gradual decay (Fig. \ref{fig:time}(a)). Zooms highlight three phases: excitation, oscillation with peak amplitude and residual vibration after ring down (left to right). Figures \ref{fig:time}(b)--\ref{fig:time}(d) visualize the corresponding cantilever motion. Immediately after excitation ($t {\sim} \SI{5}{\micro\second}$, Fig. \ref{fig:time}(b)), the cantilever first vibrated in the third longitudinal mode (L3). At a time around $t {\sim} \SI{70}{\micro\second}$ (Fig. \ref{fig:time}(c)), peak vibration amplitude was reached and we found the cantilever to oscillate simultaneously in several longitudinal and torsional modes, as can be appreciated by the minima (blue) and maxima (red) moving alongside and sideways on the cantilever beam in a wobbly fashion. After ring down ($t {\sim} 1 ms$, Fig. \ref{fig:time}(d)), the cantilever mainly vibrated in the second-order longitudinal mode (L2) with a single-digit nanometer amplitude. Visual inspection of the time trace in Fig. \ref{fig:time}(a) indicates sub-nanometer vertical sensitivity and the clarity of the height profiles in Figs. \ref{fig:time}(b)--\ref{fig:time}(d) show that our method is capable of providing reliable time-resolved phase imaging at 100 ns time resolution.

\begin{figure}[tb]
		\centering
		\includegraphics[width=0.65\linewidth, trim= 1.1in 4.5in 4.2in 1.7in, clip]{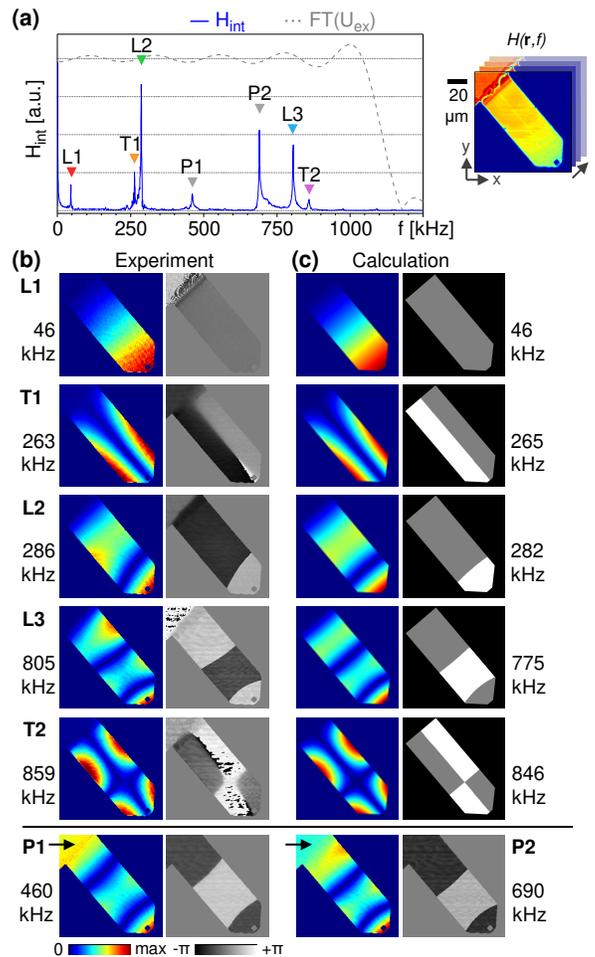}
		\caption{Frequency and mode profiles of the cantilever vibrations obtained by Fourier analysis. (a) Spatially integrated spectrum $H_\text{int}$ (b) Experimental mode profiles at the vibration frequencies identified in (a) L1-3 longitudinal modes, T1-2 torsional modes, P1-2 modes attributed to resonances in the piezoactuator (c) Calculated mode profiles of a model cantilever of \SI{100}{\micro\meter} length x \SI{37}{\micro\meter} width x \SI{0.6}{\micro\meter} height; the material was $\text{Si}_3 \text{N}_4$ with Young's modulus adjusted to match the fundamental resonance. }	
		\label{fig:freq}
	\end{figure}

By temporal Fourier transform $H(\vect{r},f)=\mathcal{F}(h(\vect{r},t))$ of the recorded surface profiles $h(\vect{r},t)$ we resolved vibration frequencies and mode profiles of all excited out-of-plane vibrations. In case of linear systems, this analysis yields a complete description of the behavior of micromechanical systems[12]. We first calculated the spatially integrated spectrum $H_\text{int} (f)=\sum_\vect{r}|H(\vect{r},f)|$  and found a distinct series of peaks in the frequency range from 0 to 1 MHz (Fig. \ref{fig:freq}(a)). By plotting the associated mode profiles $H(\vect{r},f)$ in amplitude and phase in Fig. \ref{fig:freq}(b), we identified five peaks as longitudinal flexural modes L1-L3 and torsional modes T1-T2 of the cantilever. COMSOL calculation of a model cantilever qualitatively reproduced the measured resonance frequencies and mode profiles (Fig. \ref{fig:freq}(c)). Interestingly, peaks P1 and P2 could not be assigned to any of the calculated vibration frequencies. Closer inspection of the mode profile revealed significant amplitude at the cantilever base (black arrows in Fig. \ref{fig:freq}(b), bottom), which suggested resonances in the cantilever support chip and piezoactuator - expected in this frequency range - as a possible source (see also Appendix). The multiplexed detection of vibration modes of our method provided a significant time advantage over discrete frequency methods. From the single acquisition of the 3D hologram stack, we reconstructed an effective number of ${\sim}16,000$ vibration spectra at $10,000$ FFT lines. Here we took into account the spatial resolution of \SI{0.54}{\micro\meter} in x and \SI{1.1}{\micro\meter} in y (the latter is reduced due to filtering in SOH\cite{RN175}). This, together with the near diffraction-limited resolution of the mode profiles provides a complete description of the out-of-plane vibrations in short time. Prior knowledge of the vibration frequencies is not needed, which makes our method well suited for rapid and complete analysis of out-of-plane vibration modes in micromechanical systems.

To further demonstrate the versatility of our method, we excited the cantilever with a fast, linear chirp from 20 kHz to 1 MHz in 1 ms time. Such periodic chirp is a widely used excitation waveform in spectroscopic vibration imaging of linear or weakly nonlinear devices.  Fourier analysis of the data revealed qualitatively similar vibration spectra and modes (see Appendix).

\section{Discussion}
\begin{figure}[tb]
		\centering
		\includegraphics[width=0.65\linewidth, trim= 1.1in 5.5in 4.2in 1.7in, clip]{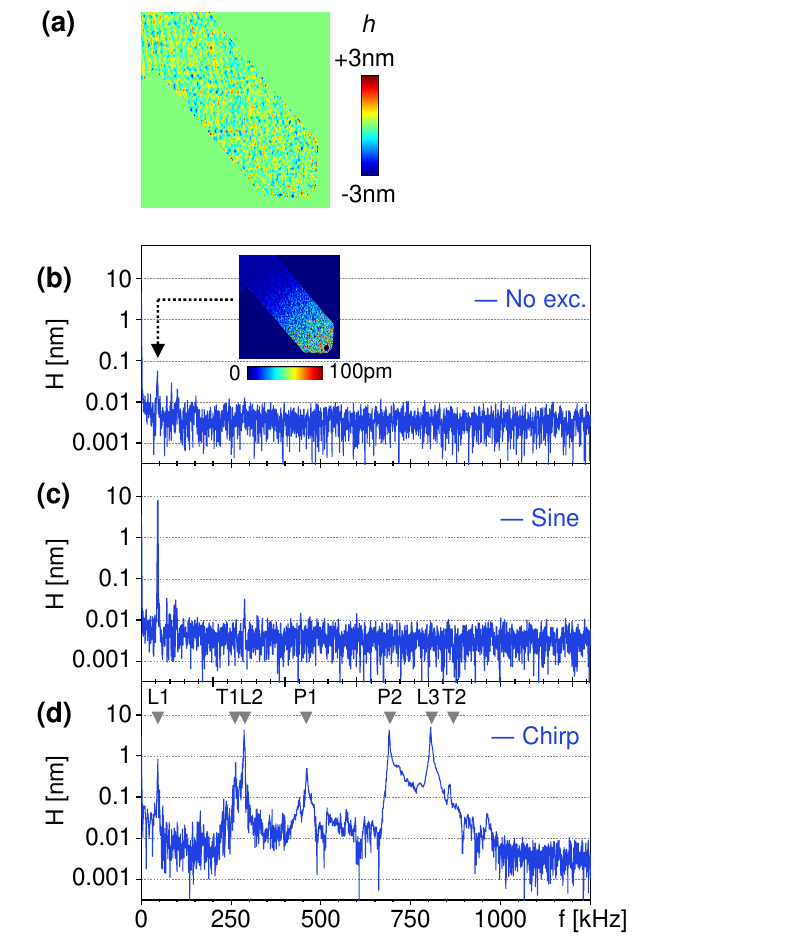}

		\caption{Vertical sensitivity provided by the presented implementation of time-resolved phase imaging. (a) Momentary surface profile $h(\vect{r},t)$ of the resting cantilever revealing 0.6 nm RMS spatial noise at 100 ns time resolution. (b-d) Vibration spectra taken at a single pixel near the cantilever tip, showing average vibration amplitude $H$. Note the logarithmic scale.(b) Resting cantilever without excitation signal applied, revealing a noise floor of $5.1 \text{pm}$ RMS, corresponding to  $\SI{230}{\femto\meter}/\sqrt{Hz}$ vertical sensitivity for the detection of vibration modes. Inset: residual L1 mode observed at 46.5 kHz. (c) Sinusoidal excitation signal tuned to the frequency of the L1 mode and at $\SI{100}{\milli\volt}_\text{pp}$ amplitude (d) Chirped excitation signal, \SI{20}{\kilo\hertz} to \SI{1}{\mega\hertz} in \SI{1}{\milli\second}, at  $\SI{500}{\milli\volt}_\text{pp}$ amplitude.}

		\label{fig:sensitivity}
	\end{figure}

We determine the vertical sensitivity of our setup by imaging the resting cantilever without any excitation signal applied. The momentary surface profile showed \SI{0.6}{\nano\meter} RMS spatial noise at the employed 100 ns integration time (Fig. \ref{fig:sensitivity}(a)). By extracting the vibration spectrum at a single pixel near the cantilever tip, we determined the noise floor in our setup to be ${\sim} 5 \text{pm}$ RMS (Fig. \ref{fig:sensitivity}(b), from 20 kHz to 1 MHz, 2 msec pixel dwell time, Hann window), corresponding to $230 \text{fm}/\sqrt{Hz}$ (Fig. \ref{fig:sensitivity}(b)). A residual L1 mode with about 75 pm vibration amplitude could thus be clearly resolved (see inset). We note that laser power, microscope and signal recording were not optimized for vertical sensitivity and expect that single-digit femtometer sensitivity could be reached with optimized interferometer setups. We further note that the employed sampling rate of 10 MHz yielded an effective 10x oversampling of the optical phase signal with respect to the bandwidth of the excitation pulse. According to the Nyquist criterium, 2 MHz sampling rate would have already provided unambiguous sampling, however, sampling at a higher rate equal the bandwidth of the optical detector (10 MHz) provided a noise advantage because noise is prevented from folding back to the signal range\cite{RN216}.

We briefly compare the detection sensitivity provided by discrete-frequency (sinusodial) and broadband (chirped) sample excitation. Sinusoidal waveform excitation tuned to the L1 vibration mode yielded a vibration spectrum exhibiting a single peak at $\SI{46}{\kilo\hertz}$ on a largely homogeneous noise floor, as expected (Fig. \ref{fig:sensitivity}(c)). Chirped waveform excitation in the frequency range of 20 kHz to 1 MHz (see Appendix) revealed all of the previously observed vibration peaks (Fig. \ref{fig:sensitivity}(d)). The noise floor in both modalities is similar to that of the unexcited cantilever, as it is mainly determined by the noise of the interferometric read-out of the surface vibrations, i.e. light-induced noise (shot noise), detector noise (thermal noise of the detector/preamplifier), and environmental noise (e.g. acoustic vibrations in the interferometer, air turbulence) \cite{RN194}. We note that residual vibration signal was observed between the vibration peaks for the chirped data set, which we attribute to weak modes belonging to the cantilever-piezo system, also observed previously in similar configurations \cite{RN143,RN149,RN160}. Discrete-frequency methods for vibration imaging provide more efficient excitation of individual vibration modes, as can be recognized by comparison of the L1 peak amplitude in Figs. \ref{fig:sensitivity}(c)--\ref{fig:sensitivity}(d) ($\SI{9.3}{\nano\meter}$ at $\SI{100}{\milli\volt}_\text{pp}$ sinusoidal excitation vs. $\SI{1.0}{\nano\meter}$ at $\SI{500}{\milli\volt}_\text{pp}$ chirped excitation), and thus might be more suitable for imaging weak vibration modes. However, spectral imaging is time consuming as repeated imaging is required for each individual vibration frequency. With the presented cantilever sample, the broadband chirped waveform yielded sufficiently strong excitation of all out-of-plane vibrations with amplitudes well above the noise floor of the instrument. At the chosen pixel, the obtained signal-to-noise ratio was in the range of 40:1 (T2) up to 1000:1 (L3). In this and similar cases, frequency-multiplexed vibration imaging is expected to resolve all excited vibration modes with sufficient signal-to-noise ratio. At the same time, the need for only a single data acquisition provides a substantial reduction in imaging time over discrete frequency approaches.
 
Temporal demultiplexing of the detector signal $U_\text{det}(T)$ (Fig. \ref{fig:setup}) separates phase retrieval and time resolution into the spatial and temporal domain, respectively. More precisely, a synthesized spatial carrier frequency  $k_\pll$  encodes the phase across the image during the scanning process by reflection of the reference beam from the slowly moving reference mirror. This effectively yields a quasi-homodyne measurement at each pixel, and particularly LOs for phase encoding pertaining to heterodyne interferometry are not needed. Thus, the large detection bandwidth of homodyne interferometry can be exploited which is only limited by the photodetector response time and bandwidth of the signal processing electronics\cite{RN214}. In comparison, heterodyne interferometry encodes phase information with a high-frequency LO in the temporal domain. Although vibration frequencies above the LO can be accessed, the highest achievable bandwidth is limited by the LO\cite{RN213}, and consequently, time resolution is theoretically limited to the order of $2-30\SI{}{\nano\meter}$ for typical LO frequencies in the range of 35 - 613 MHz. Furthermore, special Bragg cell designs are required for the generation of high-frequency LOs in the sub-GHz regime and at sufficient conversion efficiency\cite{RN213}, as well as synchronization of the LO to data acquisition.

In the current implementation of our method time resolution was limited \SI{100}{\nano\second} owing to the use of standard photodetector technology. We anticipate that our method can provide picosecond temporal resolution if the fastest available photodetector (<10 ps rise time) and digitizer technology (>100 GS sampling rate) is used. Recording of 3D hologram stacks at such scaled-up digitizer rates is not limited by the PZM velocity or the stage scanning velocity as a result of the separation of phase retrieval from time resolution. Faster digitizer rates provide finer time resolution in the form of an increased number of holograms in the 3D hologram stack, while the spatial carrier frequency $k_\pll$ in each of these holograms is independent of the digitizer rate and only determined by the PZM and stage scanning velocity\cite{RN176, RN175}. Note that PZM velocity or the stage scanning velocity may become a limiting factor if faster acquisition of the entire 3D hologram stack (i.e. faster imaging) is desired.

\section{Conclusion}

We have developed a new modality for transient vibration imaging of micromechanical systems by implementing synthetic optical holography in confocal microscopy. We have applied it to highly sensitive, frequency-multiplexed vibration imaging based on broadband sample excitation with sinc and chirped pulses and subsequent Fourier analysis. Owing to the multiplex advantage of our method, we simultaneously obtained vibration mode profiles (256 x 256 pixels) at $10,000$ FFT lines from a single data set acquired in only 4.2 minutes. With an AFM cantilever as test sample, we so obtained a complete characterization of all out-of-plane in the spectral range between 0 and 1 MHz and with $230 \text{fm}/\sqrt{Hz}$ vertical sensitivity. Further, the holographic approach of our method could provide a larger bandwidth than heterodyne interferometry, which is in principle only limited by the bandwidth of the photo detection and signal recording. We envision transient vibration imaging with picosecond time resolution as well as rapid, automated characterization MHz and GHz out-of-plane vibration modes in micromechanical systems as potential applications.

\section*{Appendix A: Chirped waveform excitation}
\begin{figure}[h]
		\centering
		\includegraphics[width=0.65\linewidth, trim= 1.7in 3.2in 2.9in 2.7in, clip]{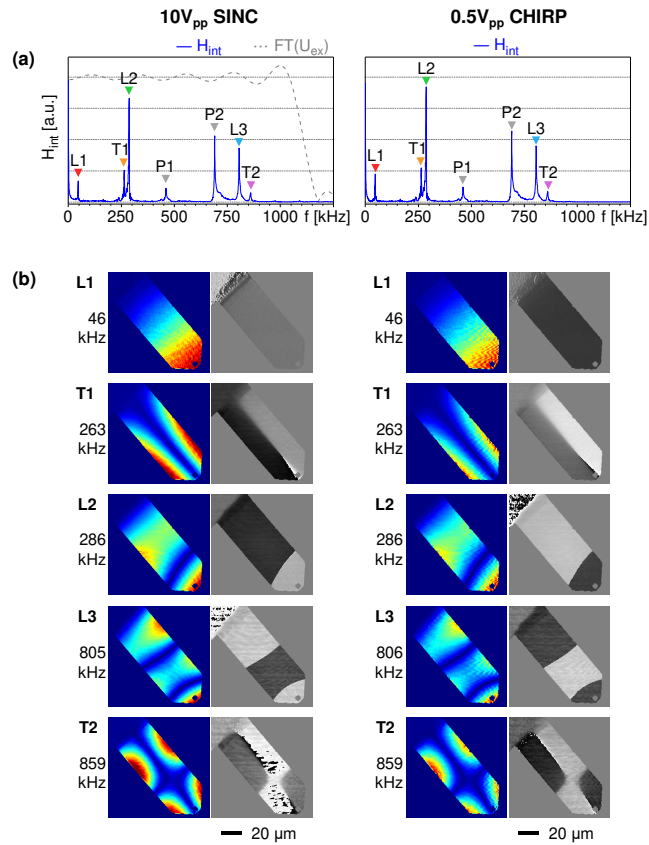}
		\caption{Comparison between sinc and chirped excitation waveforms. (a) Spatially integrated spectrum $H_\text{int}$ (b) Experimental mode profiles at the vibration frequencies identified in (a).}	
		\label{fig:sinc_chirp}
	\end{figure}

Periodic chirp is a widely used excitation waveform in spectroscopic vibration imaging of linear or weakly nonlinear devices. To demonstrate that this possibility is supported by our method, we excited the cantilever with a fast, linear chirp from 20 kHz to 1 MHz in 1 ms time. Fourier analysis of the data revealed qualitatively similar vibration spectra and modes, as comparison with the sinc-waveform data set  shows (Fig. \ref{fig:sinc_chirp}). 
	
\begin{figure}[p]
		\centering
		\includegraphics[width=0.65\linewidth, trim= 1.1in 5.2in 4.2in 1.7in, clip]{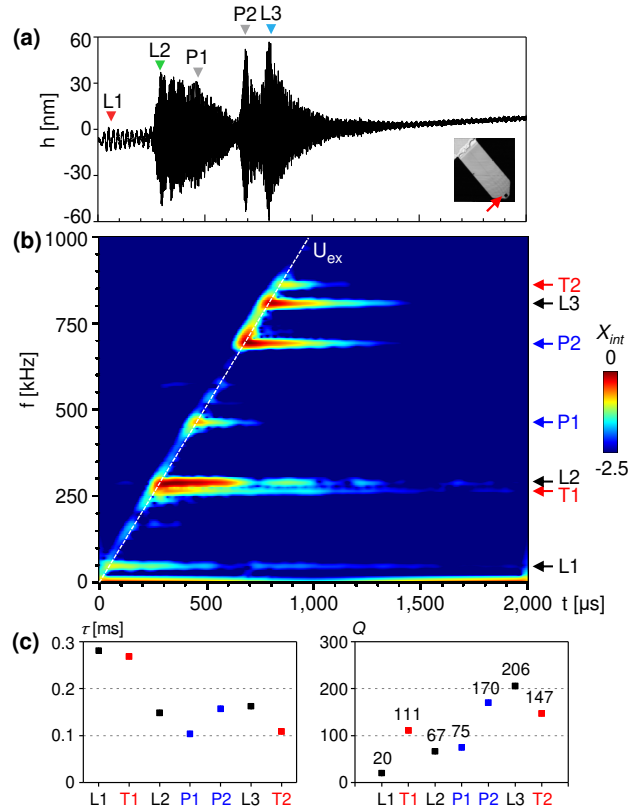}
		\caption{Monitoring vibration excitation and decay with periodic chirped excitation. Chirped excitation waveform $U_\text{ex}$ linearly sweeping from 20 kHz to 1 MHz in 1 ms was applied. (a) Time trace of the vertical position $h(t)$ extracted at a pixel near the cantilever tip. (b) Spatially integrated spectrogram $X_\text{int} (\tau,\omega)$ of vibrations excited in the cantilever as obtained by Short-Fourier Transforms. (c) Decay time $\tau$ and quality factor $Q$ of the vibration modes extracted from (b). \textcolor{urlblue}{Visualization 2} shows a movie of the cantilever vibrations.}	
		\label{fig:chirp_stft}
	\end{figure}

Additionally, we show that mode excitation and decay can be visualized with Short-time Fourier Transforms (STFT). To this end, we calculated the spatially-integrated spectrogram according to  $X_\text{int} (\tau,\omega)=\sum_\vect{r}|\text{STFT}(h(\vect{r},f))|$ from the chirped data set $h(\vect{r},t)$ using a Gaussian window of \SI{50}{\micro\meter} width. As the excitation frequency $U_\text{ex}$ was quickly swept, mechanical vibrations were excited in the cantilever beam one after another. This can be recognized with the time trace of $h(\vect{r},t)$ extracted near the tip (Fig. \ref{fig:chirp_stft}(a)), as well as by the appearance of horizontal lines on the right side of the dashed line ($U_\text{ex}$) in the spectrogram (Fig. \ref{fig:chirp_stft}(b)). We determined the decay time $\tau$ and quality factor $Q$ of the vibration modes by exponential fits along these lines (Fig. \ref{fig:chirp_stft}(c)), revealing the general trend of higher Q-factors for higher-order resonances. We attribute this finding to a reduced viscous damping as the experiment was performed at ambient temperature and pressure.

\section*{Appendix B: Investigatin of vibration modes P1, P2}
\begin{figure}[p]
		\centering
		\includegraphics[width=0.85\linewidth, trim= 1.1in 8.0in 2.6in 1.5in, clip]{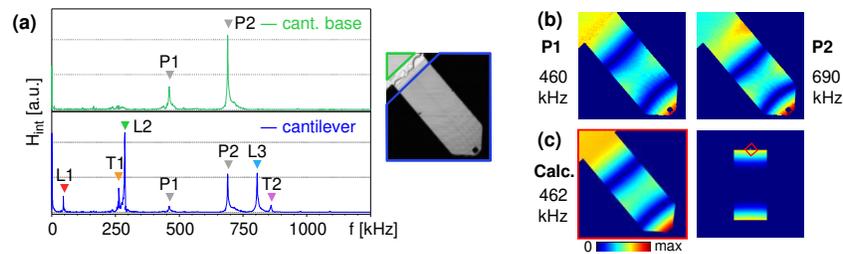}
		\caption{Investigation of vibration modes P1 and P2. (a) Spatially integrated spectrum $H_\text{int}$ over the cantilever base and the cantilever (areas marked in the respective color in the optical image). Data was scaled to fit. (b) Experimentally observed mode profiles P1 and P2. (c) COMSOL simulation of cantilever on support chip, showing a flexural mode of the support chip at $\SI{462}{\kilo\hertz}$ (left panel shows zoom on cantilever).}
		\label{fig:piezo_reso}
	\end{figure}

Spatially selective integration of the data set shown in Fig. \ref{fig:freq} revealed more insight into the nature of the modes P1 and P2 (Fig. \ref{fig:piezo_reso}). While all of the previously observed modes were observed on the cantilever itself (blue area), only modes P1 and P2 were observed at the cantilever base (green area, showing part of the cantilever support chip). COMSOL simulations of the cantilever support chip (3.4 mm x 1.6 mm x 0.5 mm, made of pyrex-glass, assuming fixed constraint on the bottom surface (2.0 mm x 1.6 mm) to account for mounting on the piezo actuator) indeed revealed a flexural mode at $\SI{462}{\kilo\hertz}$. The corresponding mode profile predicted significant vibration amplitude at the cantilever base as well as weak excitation of the L3 mode (Fig. \ref{fig:piezo_reso}(c)), as observed in the experiment  (Fig. \ref{fig:piezo_reso}(b)). This suggests that mode P1 ($\SI{460}{\kilo\hertz}$) is likely caused by a resonance mode of the support chip. The model COMSOL simulations did not reveal any resonance near P2 ($\SI{690}{\kilo\hertz}$), which suggests that P2 might be caused by resonances in the piezo actuator expected in this frequency range (Model PD050.31 from Physik Instrumente (PI) GmbH, specified unloaded resonance frequency $\SI{>500}{\kilo\hertz}$ ). 

\section*{Appendix C: Media}

\textcolor{urlblue}{Visualization 1} shows a movie of transient vibrations in an AFM cantilever excited with a sinc pulse (0 to 1 MHz bandwidth) with 100 ns time resolution. \textcolor{urlblue}{Visualization 2} shows a movie of transient vibrations in an AFM cantilever excited with a linear chirp pulse (20 kHz to 1 MHz in 1ms) with 100 ns time resolution.

\section*{Funding}
Spanish Ministry of Economy, Industry, and Competitiveness (MAT2015-65525-R); Spanish Ministry of Economy, Industry, and Competitiveness, Maria de Maeztu Units of Excellence Programme (MDM-2016-0618); European Union Horizon 2020 research and innovation programme, Marie Skłodowska-Curie Actions (H2020-MSCA-IF-2014 655888).

\section*{Disclosures}
MS, PSC, RH are authors of US patent 9,213,313.



\bibliography{microscopy}

\end{document}